\documentclass[12pt,a4paper]{article}
\usepackage{jheppub}
\usepackage{tikz}
\usepackage[FIGTOPCAP]{subfigure}
\usepackage{amsmath}
\usepackage{amsfonts}
\usepackage{slashed}
\usepackage{amssymb}
\usepackage{tikz}
\usetikzlibrary{calc}
\usepackage{tensor}
\usepackage{amsmath}
\usepackage{amssymb}
\usepackage{amsthm}
\usepackage{psfrag}
\usepackage{graphicx}
\usepackage{color}
\usepackage{bbm}
\usepackage{color}

\newcommand*{\boxcolor}{orange}
\makeatletter
\renewcommand{\boxed}[1]{\textcolor{\boxcolor}{%
\tikz[baseline={([yshift=-1ex]current bounding box.center)}] \node [rectangle, minimum width=1ex,rounded corners,draw] {\normalcolor\m@th$\displaystyle#1$};}}
 \makeatother

\usepackage{hyperref}

\newcommand{\exclude}[1]{}

\newcommand{\beq}{\begin{equation}}
\newcommand{\eeq}{\end{equation}}
\newcommand{\bea}{\begin{eqnarray}}
\newcommand{\eea}{\end{eqnarray}}
\long\def\/*#1*/{}

\newcommand{\p}{\partial}

\newcommand{\Tr}{{\rm Tr}}
\newcommand{\OO}{{\cal O}}

\newcommand{\vev}[1]{\left\langle #1\right\rangle}

\newcommand{\junk}[1]{}

\renewcommand{\Re}{\mathrm{Re}}

\title{\center{On holographic disorder-driven \\metal-insulator transitions}}
\author[a]{Matteo Baggioli \note{mbaggioli@ifae.es}}
\author[a]{, Oriol Pujol{\`a}s \note{pujolas@ifae.es}}
\affiliation[a]{Institut de F\'{i}sica d'Altes Energies (IFAE), Universitat Aut\`{o}noma de Barcelona,\\ The Barcelona Institute of
Science and Technology, Campus UAB,\\ 08193 Bellaterra (Barcelona), Spain}

\abstract{We give a minimal holographic model of a disorder-driven metal-insulator transition. It consists in a CFT with a charge sector and a translation-breaking sector that interact in the most generic way allowed by the symmetries and by dynamical consistency. In the gravity dual, it reduces to a Massive Gravity - Maxwell model with new direct couplings between the Maxwell and metric that are allowed when gravity is massive. 
We show that, generically, the effect of disorder is to decrease the DC electrical conductivity. This happens to such an extent that the conductivity does not obey any lower bound and can  be very small in the insulating phase.
In some cases, the large disorder limit produces gradient instabilities that hint at the formation of modulated phases.  
 
}

\begin{document}

\maketitle
\color{black}

\section{Introduction}
\subsection{Disorder and Holographic Massive Gravity}

Disorder in condensed matter physics   (see e.g. \cite{zimanbook,disrev,mottrev,Cardy,AltlandSimons} for reviews) is a very interesting problem. Known also as quenched or static disorder, it refers to the lack of underlying spatial order and therefore it applies to a large set of materials, ranging from amorphous solids to real-world crystals where impurities and deviations from an otherwise perfectly regular lattice structure are often inevitable.
It is also clear from the start that one may distinguish various forms of disorder:  there are many ways how impurities can be introduced and so how they impact on low energy physical quantities such as the electronic conductivity. Physical phenomena that are consequences or manifestations of disorder include Anderson localization and a some types of metal-insulator transitions, whose theoretical description which is still being debated \cite{MottInsul2,Imada:1998zz,MottInsul}.

From the  Quantum Field Theory (QFT) perspective, disorder poses a quite peculiar problem.
Physically, in the effective QFT description of the conduction quasi-electrons, it is clear that disorder can be introduced as some kind of mass term (or coupling constant) that has random and {\em spatial dependence}.
In more abstract QFT terms, this can be introduced as a deformation of the Lagrangian of the form $\delta {\cal L} = -m(x){\cal O}$ with  ${\cal O}$ {\em e.g.} a fermion bilinear and $m(x)$ a random function of space.
Aside from the obvious complication that one needs to average somehow over the realizations of $m(x)$, there are two immediate  implications of this. 
First, extracting the physical consequences of this kind of deformation can be a difficult task especially towards the low energy properties of the material. 
This can be understood from  QFT reasoning because this type of   deformation is often a {relevant} one, in the Renormalization Group (RG) sense.\footnote{For instance, a randomly varying mass term $m(x)$  in D+1 dimensions is an operator of effective dimension $3D/2$, because $\vev{m(x) m(y)}=\delta^D(x-y)$.
Thus, for a free fermion its effects grow towards low energies for $D\leq2$ (in  $D=2$ it often becomes a marginally relevant, see {\em e.g.} \cite{AltlandSimons}). It is not obvious how this counting generalizes to strongly coupled theories (possibly with no weakly coupled quasi-electron operators). Below we argue that the relevance of the `disorder couplings' in strongly coupled CFTs might be actually even more generic than  suggested by this counting.}
Quite generally, then, disorder can have a strong impact on the low-energy properties of materials (such as transport properties). In this sense, disorder often defines a non-perturbative problem.  
Second, the randomly space-dependent mass $m(x)$ picture already makes manifest that the most essential feature of disorder has to do with the (complete) breaking of space-translation invariance and consequently with momentum dissipation. Indeed, in a perfect crystal (that is, at zero disorder) and zero temperature $T$, translations are broken only partially, down to a discrete subgroup. The Bloch quasimomentum then is exactly conserved and this is enough to prevent the full relaxation of momentum at low energies. Therefore, momentum relaxation is ultimately linked to disorder (at $T=0$), that is to a complete breaking of translation invariance\footnote{\color{black} 
Note that in principle in a perfect lattice  ({\em i.e.}, with no disorder) with appropriate umklapp-like  interactions one can also have full momentum relaxation, see e.g.  \cite{Hartnoll:2012rj,Donos:2012js} -- we thank Sean Hartnoll for pointing this out. In any case, what really matters for the interpretation that we propose below of our results is that  disorder leads to momentum relaxation, which is certainly robust. 
\color{black}}. The strength of this disorder and how it depends on the scale can be formalized  by stating that the theory contains (is coupled to) a {\em translation breaking (TB) sector}, the operational meaning of which will become more clear below. \\

Holographic massive gravity (HMG) provides a somewhat unexpected and new approach to tackle this problem. HMG, here, stands for massive gravity in AdS-like spacetimes, which allow for a straightforward interpretation of the physics in terms of CFT deformations by way of the holographic dictionary. 
HMG then enjoys two very good reasons to be related to disorder: it concerns broken translation invariance and it implements non-perturbative RG flows.



Indeed, the parallel between HMG and disorder can be seen in many ways. 
In the EFT description, disorder leads to momentum non-conservation, so the  conservation of the stress-tensor $T_{\mu\nu}$ is  {\em modified} (see {\em e.g.} \cite{Davison:2013jba}). 
Upon the gauge/gravity duality, the CFT $T_{\mu\nu}$ is carried by the metric, so it unsurprising that this is linked to  {\em modified gravity} theories such as massive gravity, where the dynamical behaviour of gravity is modified at low energies. 
There is a good match between the types  (solid/fluid) of  massive gravity theories and  types of possible disordered materials (solids/fluids) has been recently worked out \cite{Alberte:2015isw,Alberte:2016xja}. 
The most important addition that  massive or modified gravity models give at the level of local and dynamical degrees of freedom is the set of Stuckelberg fields -- the longitudinal polarizations of the graviton. Similarly, the low energy EFT of real solids and fluids incorporate an additional dynamical TB sector. As is very well understood \cite{Leutwyler:1993gf,Leutwyler:1996er}, this sector  includes  light local degrees of freedom, the phonons, in the limit when the breaking is spontaneous. 
In the holographic models, then, the set of Stueckelberg fields are dual to the TB sector and in particular contain the phonon degrees of freedom. 
The appearance of light transverse phonon modes in  `solid' HMG models with small degree of explicit breaking ({\em weak disorder}) has been indeed confirmed recently  \cite{Baggioli:2014roa}\footnote{The TB sector can break translations {\em spontaneously} and/or {\em explicitly}. Gapless phonons are the manifestation of a puerly spontaneous breaking. 
At the same time, phonons are known to be present and exactly massless in a perfect crystal, {\em i.e.}, at zero disorder. 
It is natural, then, to identify disorder as the {\em explicit} breaking of translations \cite{Baggioli:2014roa}, and that the phonons obtain a mass-gap in presence of it. }.

%
%

Notice that the `holographic' adjective in HMG also plays a central role in the following discusion. First,  because  it provides a non-perturbative definition of the RG flow that still gives computational control (even with critical or quasi-critical behaviour). Second, because  it allows to include  dissipative, thermal and finite-density effects in a quite straightforward way by studying black brane solutions. Of course, attention must be paid to the possible artifacts introduced by taking the extreme case that the field theory are close to having a gravity dual. For the moment, though, it seems that this clearly gives more benefits than drawbacks.  \\

Following this logic, then, let us try to understand more closely how exactly is disorder encoded in the HMG models. The HMG asymptotically AdS planar black brane solutions can be interpreted as certain deformations of CFTs that contain a TB sector -- the CFT operators dual to the Stueckelberg fields. Since these are shift-symmetric in the bulk, they correspond to {\em exactly marginal} operators of the CFT -- let us denote them by $\OO^I$ with $I$ running over the numer of space dimensions\footnote{Let us emphasize that just like in massive gravity theories, one can describe the same physics in terms of explicit additional degrees of freedom (the Stueckelberg fields) or directly in terms of a metric (in the `unitary gauge') with modified dynamics. The difference in the gravity side is just a gauge choice. In the dual CFT with momentum dissipation, therefore, one can also have two equivalent descriptions, in terms of a set of TB operators $O^I$ or in terms of modified dynamics for the stress tensor operator $T_{\mu\nu}$. Either way, the CFTs include a $T_{\mu\nu}$ with modified local conservation. The operational meaning of why and how these CFTs include disorder seems more clear in terms of the TB sector dual to the Stueckelberg fields. Perhaps there is an equivalent formulation directly in terms of the  $T_{\mu\nu}$ alone. }. These are dynamical operators (degrees of freedom) in the CFT and so they are also equipped with their conjugated sources -- call them $S^I$. Inspection of these HMG backgrounds reveals that the external sources $S^i$ take a linear dependence in the boundary spatial coordinates, $S^I = \alpha \,\delta^I_j\, x^j$ (with  $\delta^I_j$ the Kronecker delta). It is clear then that translations are broken explicitly, as in a standard Lagrangian deformation. 
Dropping the distinction between internal ($I$) and space ($i$) indices, these backgrounds correspond to Lagrangian deformations of these CFTs of the form
\beq\label{disorderDef}
\delta {\cal L} = \alpha \;x^i \;\OO^i
\eeq
with $\alpha$ a `disorder coupling  constant'  that is going to acquire a running. Two important points follow from this observation. 
First, the disorder deformation $x^i \,\OO^i$ is {\em relevant} and of dimension $d-1$ (in $d$ space-time dimensions), because $x^i$ has mass dimension $-1$. It is clear, then, that disorder has a strong impact at low energies quite generally, at least for the strongly correlated materials that admit a dual gravity description.

Secondly, \eqref{disorderDef} is a {\em standard} Lagrangian deformation, that does not need any averaging over random field realizations (replica tricks and alike) and still it is strictly compatible with a completely homogeneous ground state. In HMG this is obvious because the holographic backgrounds dual to the deformations \eqref{disorderDef} are completely homogeneous\footnote{The reason why \eqref{disorderDef} is compatible with homogeneity of the ground state seems quite straightforward to understand in the `free unparticle' limit. The Lagrangian for such marginal $\OO$ in $d$ spacetime dimensions is, formally, $\OO (\Box)^{-d/2} \OO + 2 \OO S$. Introducing the source $S=x$ gives $\OO=(\Box)^{d/2}\,x=0$ which is indeed compatible with homogeneity.}. In hindsight, this is directly tied to the fact that the TB sector is composed of marginal operators.
The deformation \eqref{disorderDef}, then, seems to capture the averaged effect from a source of disorder that is distributed homogeneously\footnote{Note that the fact that disorder can be introduced as a standard deformation like \eqref{disorderDef} (which does not require averaging) seems in principle exportable beyond the holographic techniques. The only important requirement is that at zero disorder the theory contains (a set of) exactly marginal scalar operators.}. 

In turn, this is what makes the HMG method very convenient for computations -- it is an inherently homogeneous description of disorder.
%
%
The properties of the disorder coupling constant (how fast it grows in the IR) are model-dependent. In the holographic models below, it is quite easy to model the disorder RG flow quite directly -- it basically depends  the choice of the function $V(X)$ below. In particular, one can easily construct models where the disorder grows large `suddenly' at at low momenta (is peaked sharply near the IR part of the geometry). Since this corresponds to a mostly spontaneous breaking, these models display light phonon quasinormal modes \cite{Baggioli:2014roa}\footnote{A similar behaviour has been observed also with the spontaneous {\em v.s.} explicit breaking of scale invariance in Lorentz-invariant holographic RG  flows and the corresponding appearance of a (naturally) light dilaton pole in the spectrum of excitations \cite{Megias:2014iwa}.}.
The main focus of this paper, however, will be on how to model that the disorder affects directly the charge carrier sector. \\

The last important logical step towards modeling (semi)realistically condensed matter systems with significant effects from disorder is to take seriously that what the holographic models give is an effective description at low energies. 
In our view it is completely well-defined and motivated to take a low energy EFT-like approach that can handle systems that display IR scaling behaviour, and the holographic duality can be used exactly in this sense. 
This concept has been used almost since the beginning as a very useful   model building tool in particle physics. Recently it has been further developed and formalized into what is known as effective CFTs (see \cite{Heemskerk:2009pn,Fitzpatrick:2010zm,Fitzpatrick:2013twa} and references therein), nn part thanks to the progress in gauge/gravity duality. The main idea is that coupling the asymptotically anti-de Sitter (AAdS) dual gravity theory to some EFT defines an effective CFT with an operator content that can be mapped to the field content in the classical gravity theory. A mass gap in the (gravity + EFT) theory results in a  gap in the dimensions of the operators that participate in the dual CFT. It is not the purpose of this article to delve into under what conditions the effective CFTs are well defined or are close to having a gravity dual. We will simply use this logic as a model-building tool: since the  gravity theory is a well defined EFT and the holographic dictionary provides a clear interpretation, one can controllably build consistent models. 

\subsection{Minimal disordered holographic models}

In this note we shall concentrate on providing model of a disorder-driven metal-insulator transition as minimal as possible using these holographic tools. We shall require the following ingredients: 
\begin{enumerate}
\item we start from a UV-fixed point CFT, with operator content including: an electromagnetic current density $J_{\mu}$ and a stress tensor $T_{\mu\nu}$ with {\em modified} conservation. Equivalently, one can phrase this as including a translation breaking sector ${\cal O}^i$ (composed of the operators dual to the Stueckelberg fields $\phi^i$).

\item we consider only  having a finite density of charge carriers $\rho$ at finite temperature $T$. In our models, this corresponds to considering asymptotically AdS Reissner-Nordstrom planar massive gravity black branes.

\item we only allow ourselves to turn on one additional {\em disorder} deformation,  which implements the modified conservation of $T_{\mu\nu}$ and is incarnated in the Stuckelberg fields profile $\Phi^i\propto x^i$. This rules out the possibility of playing with additional running coupling constants (running dilatons)
\end{enumerate}

As we shall see,  minimal effective CFTs that comply with the requirements and give an MIT that is unquestionably driven by disorder do exist.

%
%
%
%
%
%

\color{black}Our results show how to overcome the obstruction recently found in \cite{Grozdanov:2015qia} to obtain disorder-driven metal-insulator transitions (MIT) in {\em simple} holographic theories. \color{black}
%
What is  meant by `simple' there relies on a number of  assumptions, the most important of which are i) there no dilaton field $\Phi$ coupled to the Maxwell sector of the form $ Z(\Phi)\,F_{\mu\nu}F^{\mu\nu}$ is allowed (similar to our condition 3 above) and ii) additional bulk fields are allowed as far as they are not charged. The authors of \cite{Grozdanov:2015qia} then argue that their conditions  basically imply that the action for the holographic dual is reduced to the form
\begin{equation}
\mathcal{S}_{bulk}\,=\,\int\,d^4x\,\sqrt{-g}\,\left({R\over2}-2\,\Lambda-\frac{F_{\mu\nu}F^{\mu\nu}}{4\,e^2}\,+\,\mathcal{V}(\phi^I)+\dots\right)
\label{ActPerm}
\end{equation}
where $\mathcal{V}(\phi^I)$ encodes the Lagrangian for a generic  neutral translation-breaking (TB) sector.
Within this restricted set of models  Ref.~\cite{Grozdanov:2015qia} then shows that
the electric DC conductivity of an holographic disordered system is bounded from below by the universal value:
\begin{equation}
\sigma \geq \frac{1}{e^2}
\label{bound}
\end{equation}
with $e$ the $U(1)$ charge that represents the unit of charge of the charge carriers.
This basically means that one can not get an insulating configuration where the electric DC conductivity at zero temperature is very small or eventually zero. As a consequence, within the models \eqref{ActPerm}, no disorder-driven MIT appears.\\

\color{black}In the present note, we shall show instead that one can certainly avoid a bound like  \eqref{bound} in minimal and natural holographic models that contain the same dynamical ingredients (operators) as well as the mutual (`electron-disorder')  interactions which are still allowed by the symmetries. Indeed, using \color{black}   EFT-like reasoning, it is clear that \eqref{ActPerm} is not the most general action allowed by the symmetries and the required field content. Clearly, there are additional couplings between the charge and TB sectors that can (and should) be included in the effective action. The  crucial new ingredient that will be relevant for the present discussion is a direct coupling between the charge and TB sectors, which we can write schematically as
\beq\label{direct}
Y[\phi^I] \, F_{\mu\nu} F^{\mu\nu}
\eeq
where $Y[\phi^I]$ stands for some function of the TB sector $\phi^I$ (or its derivatives) alone. 
Physically, even before specifying how we shall implement the TB sector and choose the $Y$ function,
it is clear that this effective interaction captures how much the TB sector affects the charge sector. 
This coupling, then, encodes the {\em charged disorder} -- the effects from ionic impurities that directly couple to the mobile charge carriers.  
From the point of view of an effective description it is all the more reasonable that this kind of disorder is encoded in a direct coupling of this form. 

The main result of this note will be that by allowing the coupling \eqref{direct} gives rise to a very reasonable disorder-like phenomenology (in a models that are completely under control) that includes a very clear disorder-driven metal-insulator transition. In particular, this implies that  absence of any generic universal lower bound like \eqref{bound} on the electrical conductivity: even in minimal purely conformal effective theories that are perfectly under control, this bound can be violated.

Let us insist that our model below complies with the criteria of \cite{Grozdanov:2015qia} regarding dynamical ingredients and symmetries. The only difference is that we allows for a possible new effective interaction that are permissible among the various CFT operators. That this new term is allowed is obvious from the effective approach to holography. It is subject to some consistency conditions, but these do not prevent it to be present. On the contrary, they force it to do interesting things, namely to reduce the conductivity at low T and produce MITs with good insulators.  We will not enter into how can this new interaction be obtained from some UV completion. We will simply limit ourselves to include this new effective interaction and analyze its consequences.

Let us also emphasize one crucial difference between our proposal and some previous models such as \cite{Cai:2015wfa,Kiritsis:2015oxa,Davison:2013txa,Gouteraux:2014hca} that use a running dilaton $\Phi$ that couples to the charge sector through a bulk term like $Z(\Phi) F^2$. These models include a new dynamical ingredient, a scalar CFT operator $\cal O$. The BB solutions are relevant deformations of the CFT by the operator $\cal O$ that already in vacuum gives rise to confinement and therefore an insulating-like behavior. In these cases, it is hard to argue that the insulating behavior is driven by disorder. In our case, instead, there is no room for doubt. There are no more dynamical ingredients in the CFT other than the TB and the charge sector, so the BB solutions represent CFTs deformed by disorder (and finite density). 
At this point we can also see that the new interaction will play a role similar to the dilaton-Maxwell coupling in the sense that the physical magnitude of the charge carried by a charge carrier  gets renormalized along the RG flow -- in our case clearly due to disorder. Furthermore, we will show that dynamical  consistency of the model requires that the renormalization is such that the conductivity is necessarily reduced at small temperatures (which is not necessarily the case for the Maxwell-dilaton coupling).

There have been other attempts to describe disorder using holography that do not make an explicit use of massive gravity \cite{ Vegh:2013sk, Blake:2013bqa,Donos:2014yya,Lucas:2014sba,O'Keeffe:2015awa,Lucas:2015lna,Garcia-Garcia:2015crx}, but either introduce explicitly position-dependent external sources or deal with the averaging  of the stochastic random potential \cite{Fujita:2008rs,Adams:2011rj,Adams:2012yi,Hartnoll:2014cua,Hartnoll:2015faa,Hartnoll:2015rza,Aharony:2015aea}.
These models are considerably less convenient to work with and they reduce upon averaging to forms of HMG \cite{Blake:2013owa}.

\color{black} Let us point out also that our models are structurally quite similar to the model presented in \cite{Donos:2012js}. Both theories contain a charge and a translation breaking sector, and a cross coupling between them. In both cases the MITs are facilitated by the cross-coupling parameter.  The main difference is that the background solution (or `ground state') in our models is homogeneous and isotropic, whereas in \cite{Donos:2012js} is homogeneous but anisotropic since  it is invariant under a helical translational symmetry. 
\color{black}\\

The rest of this note is structured as follows: in Sec.~\ref{model} we present the minimal holographic model, in Sec.~\ref{result} we discuss its electric transport properties and in Sec.~\ref{discuss} we give some discussion. Appendix \ref{cons} is devoted to the analysis of the consistency.

\section{Minimal Model}\label{model}

We consider the following model in $3+1$ dimensions\footnote{In the unitary gauge $\phi^i=x^i$, the new coupling looks like $ Y(\Tr(g^{ij}))\,F^2$, which makes manifest that once gravity is massive then  the coupling of a gauge boson to gravity can include many more terms than just the usual Maxwell term. } :
\begin{equation}
\mathcal{S}\,=\,\int d^4x\,\sqrt{-g}\,\left[{R\over2}-\Lambda-\frac{1}{4\,e^2}\,Y(X)\, F^2-\,m^2\, V(X)\right].\label{action}
\end{equation}
with $X=g^{\mu\nu}\p_\mu\phi^I\p_\nu\phi^I$ and $F^2=F_{\mu\nu}F^{\mu\nu}$. From now on we fix the charge unit to one, $e=1$, along with the normalization $Y(0)=1$ (see \cite{Baggioli:2015dwa, Baggioli:2015gsa, Baggioli:2015zoa} for previous studies and applications of similar models).
The model consists of two coupled sectors aside from the metric (which encodes the CFT stress tensor $T_{\mu\nu}$):
\begin{itemize}
\item[i)] the $U(1)$ charged sector, encoded in $A_\mu$ which corresponds to the usual charged current operator $J_\mu$ of the CFT; and 
\item[ii)] a $U(1)$-neutral translation breaking (TB) sector containing the scalar fields $\phi^I$ that (take v.e.v.s $\phi^I=\delta^I_i x^i$ and) enjoy internal shift symmetry that dictates that they enter only through derivatives, such as $X$. To a certain extent, one can think that $\phi^I$ corresponds to a set of exactly marginal operators of the dual CFT. As discussed above, there is a good amount of evidence that this kind of TB sector captures an averaged description of disorder coming from an homogeneously distributed set of impurities. 
\end{itemize}

It is quite clear from the structure of the action \eqref{action}, that the TB enters in two distinct ways, encoded in the functions $Y(X),\,V(X)$. $V(X)$ represents a \textit{neutral disorder} -- the disorder from neutral forms of impurities, that do not couple directly to the charge carriers.  Instead, $Y(X)$ captures the effects from charged impurities, that are felt directly by the  charged sector.

The way how $\phi^I$ enters in the action is of course not the most general one ($V$ and $Y$  could also depend on $\p^\mu\phi^I\p_\mu\phi^J \p^\nu\phi^I\p_\nu\phi^J$ and one could include also other tensorial contractions between $F_{\mu\nu}$ and $\p_\mu\phi^I \p_\nu\phi^I$), but we shall stick to \eqref{action} for the sake of simplicity and because it already illustrates the main differences with previous models\footnote{Let us insist that models involving holographic Q-lattices such as \cite{Ling:2015exa,Ling:2015epa} might at first seem similar to ours. However they differ in a crucial aspect: these models include charge+TB+dilaton sectors and only the dilaton  couples directly the charge sector. Therefore, these models are less minimal and it is not so clear that the MIT is disorder-driven.}.

In the language of \cite{Alberte:2015isw,Alberte:2016xja} these  models behave as (viscoelastic) solids. Similar constructions can also be done with fluid-type models where the Lagrangian instead of depending on $X$ would depend on the determinant $Z$ in the notation of \cite{Alberte:2015isw} and represent momentum-disspating fluids \cite{Alberte:2015isw}. The analysis of that case would proceed along the same lines as the one presented here, and we expect that very similar results (basically because in the background solution $Z$ also takes a nonzero value).

The model admits asymptotically AdS 
charged black brane solutions with a planar horizon topology. For arbitrary choice of $V,Y$ they take the form:
\begin{align}
&ds^2 = \frac{1}{u^2}\left[-f(u)dt^2+\frac{1}{f(u)}\,du^2+dx^2+dy^2\right]\,,\nonumber\\
&f(u)\,=-\,u^3\,\int_{u}^{u_h} \left(\frac{\rho^2}{2\, Y\left(\alpha ^2 \xi^2\right)}+\frac{m^2\, V\left(\alpha ^2 \xi^2\right)}{\xi^4}+\frac{\Lambda }{\xi^4}\right)\,d\xi\,,\nonumber\\
&\phi^I=\alpha \,\delta^I_i x^i\,\,,\,\,\,\,\,I=\{x,y\}\,,\nonumber\\
&A_t(u)\,=\,\rho\,\int_{u}^{u_h}\frac{1}{Y(\xi^2\,\alpha^2)}\,d\xi\,,
\label{ansatz}
\end{align}
where $u_h$ denotes the horizon location.

The fields $\phi^I$ are the responsible for the breaking of translational symmetry in the $\{x,y\}$ directions of the CFT (see \cite{Andrade:2013gsa} for the original work) and they are introduced in a non-linear way as proposed in \cite{Baggioli:2014roa}.

The temperature of the background geometry reads:
\begin{equation}
T\,=\,-\frac{\rho^2\,u_h^3}{8\, \pi \, Y\left(\alpha ^2\,u_h^2\right)}-\frac{m^2\, V\left(\alpha ^2 u_h^2\right)}{4\, \pi  \,u_h}-\frac{\Lambda
   }{4\, \pi  \,u_h}
\end{equation}
In the following the cosmological constant will be fixed to $\Lambda=-3$ to accomodate usual AdS asymptotics. Note we have already fixed the AdS radius $L=1$ without loss of generality.

An very important part of the present analysis concerns the conditions under which the models above are consistent -- they are free from  instabilities. We perform this analysis in the Appendix \ref{cons}. Its main outcome is that the functions $V(X), Y(X)$ that appear in the Lagrangian are subject to the constraints:
\begin{equation}
V'(X)>0\,\,\,,\,\,\,Y(X)>0\,\,\,,\,\,\,Y'(X)<0
\label{resum}
\end{equation}
Crucially, the Maxwell-Stueckelberg coupling $Y$ is allowed (and must be positive). Not only that, it must also be a decreasing function of $X$. Let us emphasize that the latter condition stems solely from the requirement that the transverse vector modes have a normal (non-ghosty) kinetic term (the actual condition is slightly less restrictive, but for simplicity we  \color{black}shall take $Y'<0$ \color{black} which is certainly sufficient and more robust). The fact \color{black}that $Y'<0$ \color{black} will have  a dramatic impact on the possibility to have a MIT.

For the purpose of this short note we shall focus on a representative `benchmark' model,
\begin{equation}
Y(X)\,=\,e^{-\kappa\,X}\,,\,\,\,\,\,V(X)=X/(2\,m^2)\,\,.
\label{themodel}
\end{equation}
This is by far not the most general model but it will suffice to illustrate the new features that can be modeled with this kind of coupling. 
Note that it suffices to take $\kappa>0$ to satisfy all the  consistency conditions. One can also anticipate that for order-one values of $\kappa$ the effects from this coupling can be rather important. 

Keeping in mind from \eqref{ansatz} that in the background solution $\bar X = \alpha^2 u^2$, in this model, the direct coupling between the  charge and the TB sectors is incarnated in $\kappa\, \alpha^2$, which can be understood as the amount of charged disorder.  Instead, the parameter $\alpha$  encodes a neutral form of disorder. From now on, then we shall refer to $\kappa$ as {\em charged disorder} and to $\alpha$ as {\em neutral disorder}.

\section{Electric response}\label{result}
Proceeding with the vector perturbations on top of the background defined in \eqref{ansatz}, one can compute numerically the optical electric conductivity and analytically its DC value. Following \cite{Amoretti:2014zha,Donos:2014cya} and skipping the details of this standard computation, we find the following analytic result for the  electric DC conductivity,
\begin{equation}
\sigma_{DC}\,=\Big[\,Y(\bar{X})+\frac{\rho ^2\, u^2}{m_{\footnotesize eff}^2} \,\Big]_{u_h}
\label{DCformula}
\end{equation}
with 
$$
m_{\footnotesize eff}^2\equiv \alpha ^2 \left(m^2\,  V'(\bar{X})-\rho ^2 \,u^4\, {Y'(\bar{X})\over{ 2\,Y^2(\bar{X}) }}\right)
\label{MMM}
$$
where $\bar{X}\,=\,u^2\,\alpha^2$ and all quantities have to be evaluated at the horizon, $u=u_h$.
\begin{figure}
\centering
\includegraphics[width=7cm]{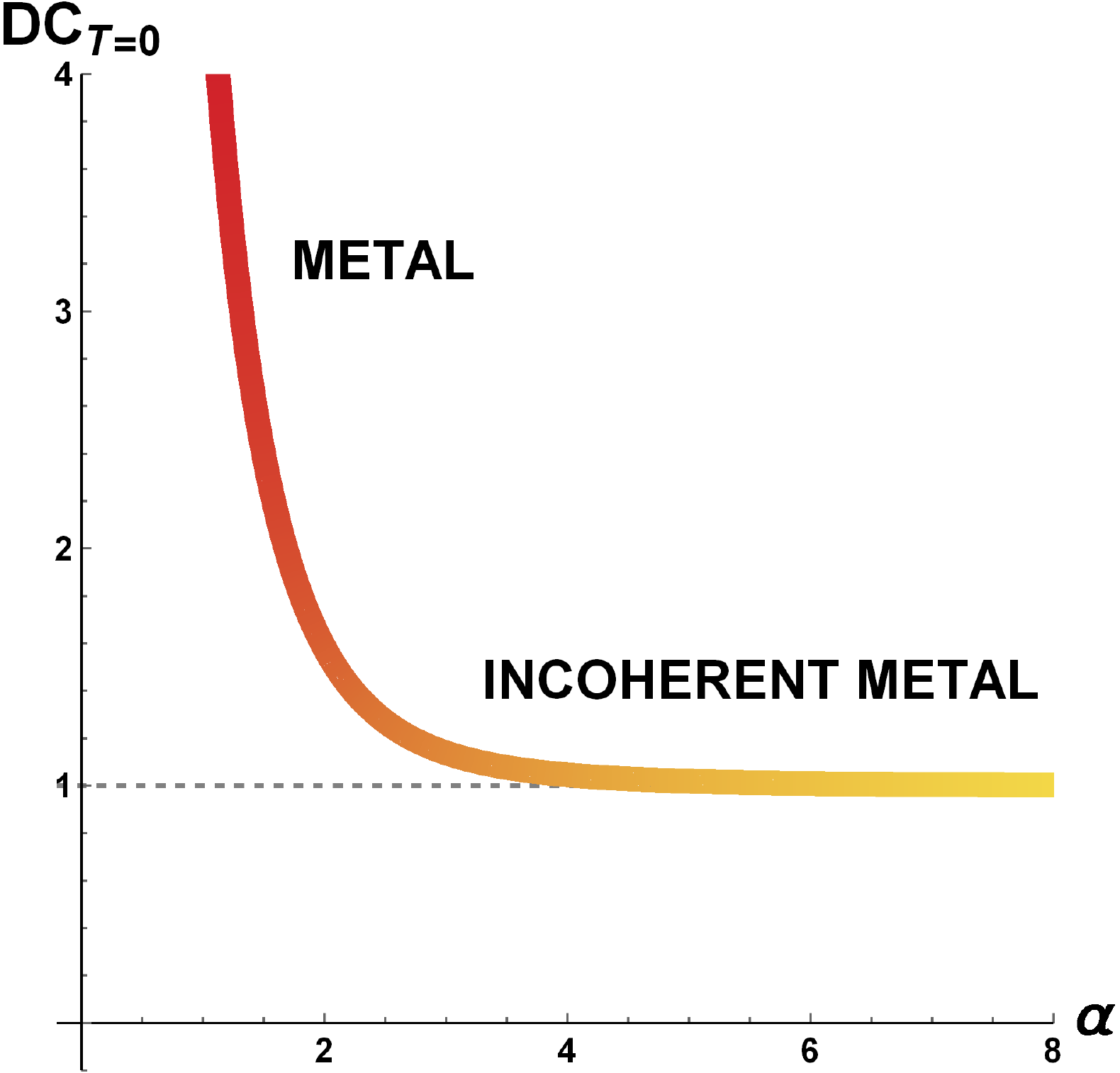}
\qquad \includegraphics[width=7cm]{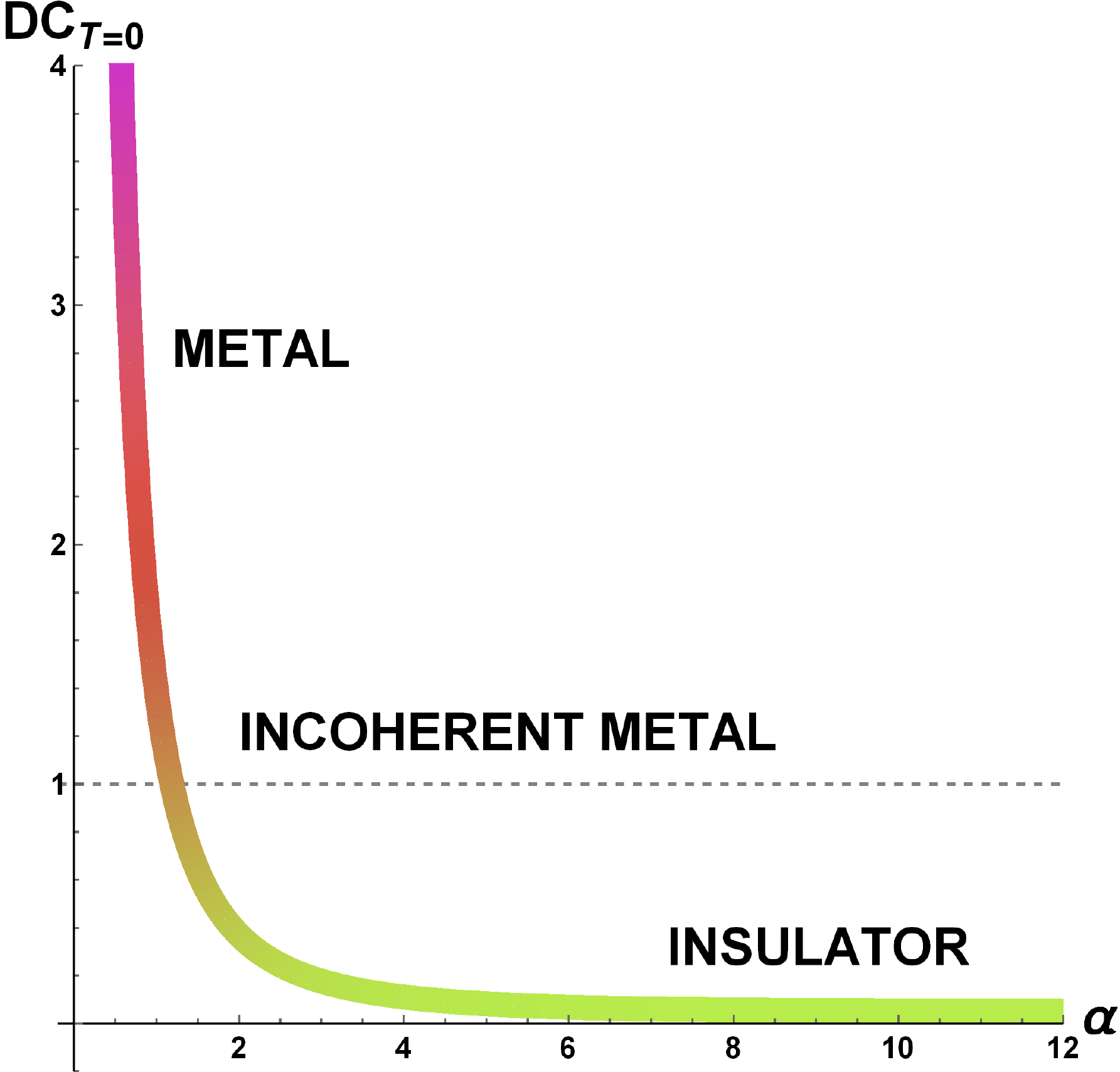}
\caption{Electric DC conductivity at zero temperature and charge density $\rho=1$ for the model \eqref{themodel} dialing the disorder strength $\alpha$ (i.e the graviton mass); \textbf{Left:} $\kappa=0$ (i.e. previous literature); \textbf{Right:} with the new coupling $\kappa=0.5$ (safe region) .}
\label{DC0}
\end{figure}
The expression \eqref{DCformula} is the main result of this work and it encodes all the interesting phenomenology which follows.
Note that one of the consistency conditions found in the Appendix \ref{cons} can be expressed as the positivity of the effective graviton mass, $m_{\footnotesize eff}^2>0$.
 
For the benchmark model \eqref{themodel}, the interesting quantities read:
\begin{align}
&T\,=\,-\frac{\rho ^2 \,u_h^3 \,e^{\kappa\,\alpha ^2 u_h^2}}{8\, \pi }-\frac{\alpha ^2 \,u_h}{8 \,\pi }+\frac{3}{4\, \pi \,u_h},\nonumber\\
&\sigma_{DC}\,=\,e^{-\kappa\,\alpha ^2 \,u_h^2}+\frac{2 \,\rho ^2 \,u_h^2}{\alpha ^2\, \left(\rho ^2 \,\kappa\,u_h^4\, e^{\kappa\,\alpha ^2 u_h^2}+1\right)}\,.
\end{align}
In this scenario we are left with only four parameters in our model: the temperature $T$, the charge density $\rho$, the neutral and charged disorder strengths $\alpha$ and $\kappa$.

The analysis of the electric DC conductivity in function of disorder is our primary task. The first interesting and new feature of the model deals with the DC conductivity at zero temperature which characterizes the nature of our '\textit{material}' (i.e. metal/insulator). In the previous massive gravity models the system could be just in a metallic phase (with a sharp Drude peak) or in an incoherent metallic phase where there is no clear and dominant localized long lived excitation. The bound of \cite{Grozdanov:2015qia} can be indeed rephrased with the statement that such a models fall down in an extremely incoherent metallic state for very strong disorder without undergoing a metal-insulator transition. Note that this is a quite unnatural behaviour in real-life experiments where usually strong disorder produces clear insulating behaviours (see \cite{MottInsul, MottInsul2} for some reviews). In our case the scenario is more complex and increasing the disorder one can reach an insulating state where $\sigma_{DC}\approx 0$ at zero temperature\footnote{\color{black}For the benchmark model considered here, at large $\kappa$ the electric conductivity saturates to a small but finite value. Certainly other choices of $Y(X)$ can lead to a conductivity that is exponentially suppressed, but this is beyond the point of the present work. 
Note  in Ref.\cite{Gouteraux:2016wxj} (appeared soon after our work) the choice $Y(X)=1-a\,X$ is made which leads to exactly vanishing conductivity at a finite value of the parameters. -- We thank Wei-Jia Li for discussions about this point. In these models, though, this immediately leads to a restriction in the parameter space in order to avoid negative $\sigma_{DC}$ which should be associated with the appearance of ghosts. \color{black}}. This is basically what we mean by disorder-driven MIT and to the best of our knowledge this model is the first holographic example of such a mechanism. The difference between our novel results and previous literature is summarized in fig.\ref{DC0}.

Another interesting feature within this model is that we can produce a metal-insulator crossover at finite temperature (as obtained in \cite{Baggioli:2014roa} ) without the need of playing with non linear potential $V$ for the scalar goldstones sector. We can in fact provide a transition between a regime with $d\sigma/dT<0$ (metallic behaviour) at large temperature to a regime with $d\sigma/dT>0$ (\textit{insulating} phase\footnote{The strict definition of an insulator a condensed matter theorist would give you is the following: $\sigma(T=0)=0$; here we are using a more realistic and phenomenological definition which is basically $d\sigma/dT>0$ . Note as one can fullfill the second requirement without accomplishing the 1st one (as in the case of \cite{Baggioli:2014roa}).}) at small temperature.

\begin{figure}
\centering
\includegraphics[width=7cm]{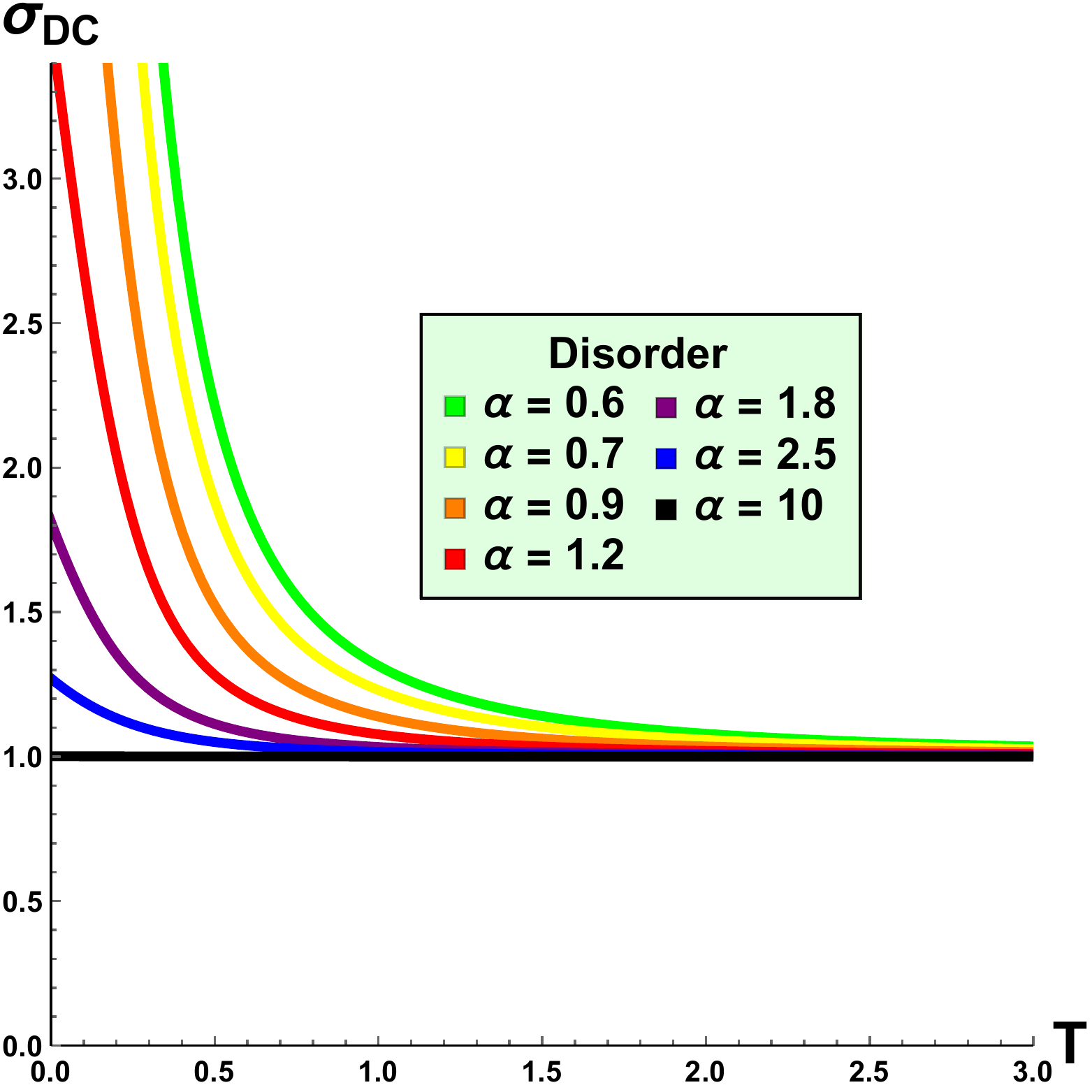}
\qquad
\includegraphics[width=7cm]{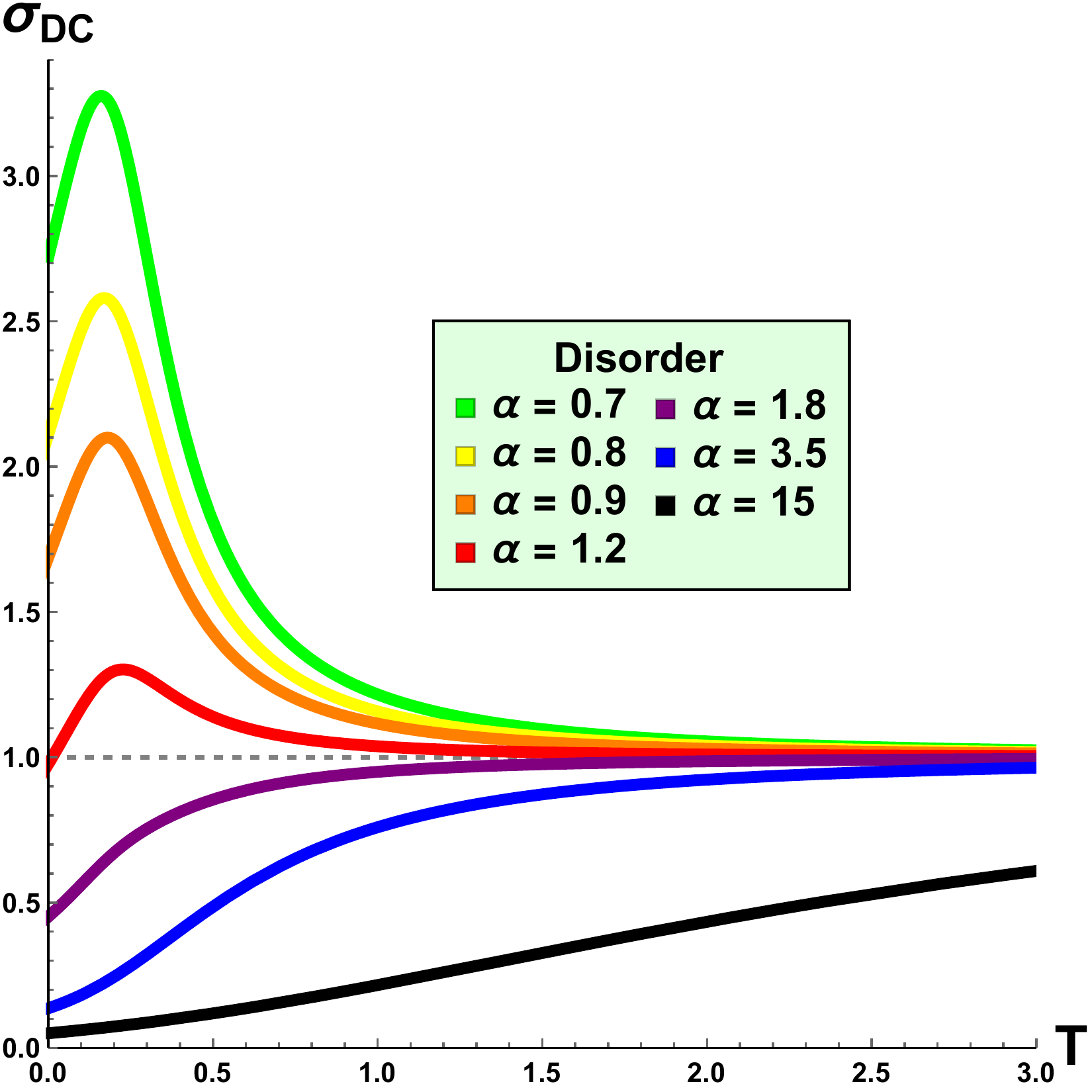}
\caption{Temperature features of the DC conductivity at different disorder strengths $\alpha$ (i.e. graviton mass) for the model considered in \eqref{themodel} with unitary charge density $\rho=1$; \textbf{Left:} metal-incoherent metal transition for $\kappa=0$ (i.e. previous literature); \textbf{Right:} metal-insulator transition for $\kappa=0.5$ (safe region).}
\label{MITplot}
\end{figure}

Most of the results of this short note are summarized in fig.\ref{MITplot} where the presence of a metal-insulator transition (in contrast with previous massive gravity models) is made evident . The picture again emphasizes how we can overcome the bound \eqref{bound} proposed in \cite{Grozdanov:2015qia} (dashed line) increasing the disorder in the system and exploiting the new parameter $\kappa$. It is definitely interesting to understand more the phenomenology of this transition and its generic features. 

\section{Discussion}\label{discuss}

In the framework of bottom-up holographic theories for condensed matter we provide a minimal example of a disorder-driven metal-insulator transition (DDMIT). By `minimal' we mean that the CFT that describes the low energy physics contains only two sectors (or operators): a $U(1)$ charge current $J_\mu$ and a translation breaking sector that can be implemented as a set of translation-breaking scalars or equivalently as a  stress tensor with modified conservation.\footnote{Importantly, we do not consider additional relevant deformations (such as running dilatons in the bulk) in order to isolate  transitions that are clearly disorder-driven.} To the best of our knowledge this is the first example of such a DDMIT with these restricted number of ingredients. 
By explicit construction, then, we overcome the impossibility found in  \cite{Grozdanov:2015qia} for obtaining  a DDMIT in the context of holography. The key element to achieve such a DDMIT was to include the allowed cross-couplings between the charge and TB sectors, which were not considered in \cite{Grozdanov:2015qia}. This result also  suggests that there is no universal lower-bound for the electrical conductivity at least within the holographic models.\footnote{Within strongly correlated materials there is a similar bound on the conductivity, known as the  Mott-Ioffe-Regel (MIR) bound. It seems, though, that there are known materials where this bound is violated. See also \cite{Hartnoll:2014lpa,Amoretti:2014ola,Kovtun:2014nsa, Grozdanov:2015djs} for 
other examples of potential  universal bounds that are being discussed.} 
\color{black}Despite the absence of any lower bound in the electric conductivity provided by our simple generalization of the implicit assumptions in \cite{Grozdanov:2015qia}, it should be very clear that we are not giving in a derivation of \textit{localization}, or that localization is the phenomenon that lies behind the bad conductivity of these holographic materials. 
For this one would certainly need to abandon homogeneity and study more complicated models  on the lines of \cite{Hartnoll:2014cua,Hartnoll:2015faa,Hartnoll:2015rza}\footnote{\color{black}We thank S. Grozdanov, S. Hartnoll, K. Schalm, A. Lucas for dissuasions on this point.}. Still, the point that we tried to illustrate is that the effective holographic models discussed here i) are consistent with the interpretation that they account in a homogeneous way for the effects of disorder,\footnote{\color{black} Note that following the same effective field theory logic it is also possible to account for self-interactions in the charge sector by introducing gauge field nonlinearities in the gravity side  \cite{Baggioli:2016oju}. \color{black}} and ii) these models can therefore provide  an effective and useful (homogeneous) description of the disordered system. This can conceivably operate in a similar same way that the Chiral lagrangian gives an effective (and predictive) description of mesons and their interactions while it does not represent a derivation of confinement neither of what is the mechanism for chiral symmetry breaking.   \color{black}
In the process, we have made an effort give a comprehensive interpretation the Holographic Massive Gravity (HMG) solutions in terms of standard Quantum Field Theory. We have argued that HMG gives a new and convenient way to tackle the problem of quenched disorder in strongly correlated materials.  

In the gravity dual, our model reduces to a theory of massive gravity coupled to a Maxwell field with a `direct' interactions between metric and charge sector in the form of a potential term for the metric that couples to the electromagnetic field strength. Equivalently, in Stueckelberg language, we include a kinetic mixing between the Maxwell and Stueckelberg fields.
We have shown that in the limit of large graviton mass (strong disorder) the model displays a metal-insulator transition and it can  reach almost zero electric conductivity $\sigma_{DC}\approx 0 $ at zero temperature.

\begin{figure}
\centering
\includegraphics[width=17cm]{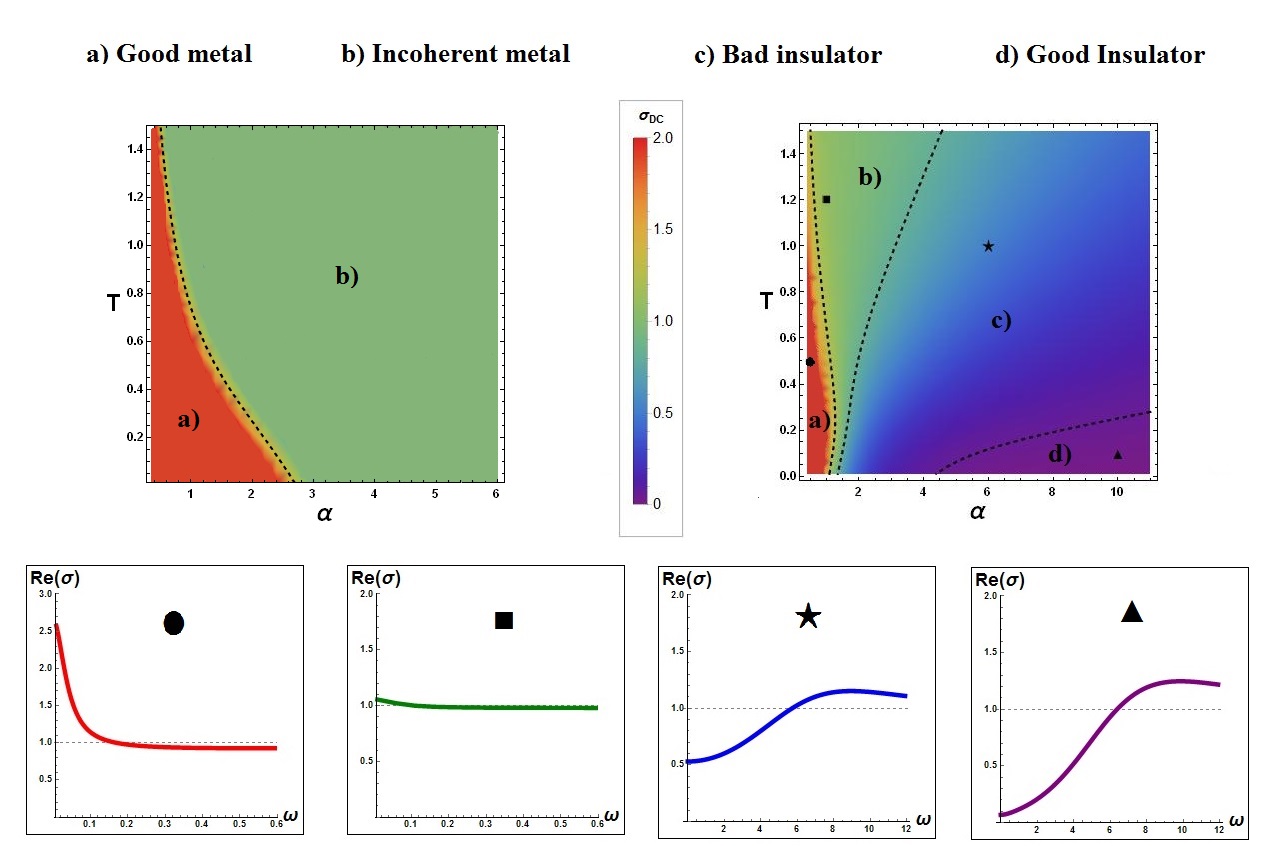}
\caption{Phase diagrams for the model \eqref{model} with unitary charge density $\rho=1$. Dashed lines correspond to $\sigma_{DC}=0.1,\,0.8\,,1.2$ and they divide the four regions: a) good metal, b) incoherent metal, c) bad insulator and d) good insulator . \textbf{Top Left:} Temperature-disorder plane with $\kappa=0$. \textbf{Top Right:} Temperature-disorder plane with graviton mass $\kappa=0.5$. \textbf{Bottom:} For every region (a,b,c,d) in the phase diagrams one representative example of $\Re\,(\sigma)$ is shown. The parameters for each one of the AC plots are pinpointed in the $T-\alpha$ phase diagram (top right) and they correspond to: $\left[\,\bullet:(\alpha=0.5,T=0.5),\,\blacksquare:(\alpha=1,T=1.2),\,\bigstar:(\alpha=6,T=1),\,\blacktriangle:(\alpha=10,T=0.2)\,\right]$ .}
\label{Phases}
\end{figure}

The main features of the model \eqref{action}  are summarized in Fig.~\ref{Phases} where we draw the phase diagrams in the temperature-disorder plane for $\kappa=0$ ({\em i.e.} previous literature) and $\kappa=0.5$ (which lies in the healthy region of the parameters space). For the known case $\kappa=0$ just metallic phases are accessible and only a crossover between metals and incoherent metals can be manifested (see {\em e.g.} \cite{Davison:2014lua, Kim:2014bza}). On the contrary, for the novel case, the phase diagram gets richer and incorporate several phases of matter depending on the parameters: good metal (a), bad or incoherent metal (b), bad insulator (c) and good insulator (d). Both the quantum phase transition (MIT) and the finite temperature crossover are present in the picture. For every phase of matter a representative example of optical conductivity is shown at the bottom of Fig.~\ref{Phases}.

Let us remark that the new cross-coupling introduced in our model \eqref{action} 
is a function $Y(X)$, whose functional form of $Y$ is constrained by a number of stability and consistency reasons (see the Appendix \ref{cons}). An interesting outcome of the consistency analysis is that if a non-trivial $Y(X)$ is introduced, it  must have a negative slope $Y'(X)<0$. In turn, this implies that the effect of $Y$ (that is, of the coupling of charge to disorder) is that conductivity has to  decrease towards low temperatures. We find this a quite powerful and generic property of these models that matches well with what one expects from disorder. 
For simplicity, we have taken \eqref{themodel} as a convenient and representative case, but our results seem to hold much more generally. 
In order to get an insulating background one just needs the function $Y(X)$ to be zero or very small for large $X$. 

We leave for future research a more thorough analysis of the optical response to search, {\em e.g.} for  the formation of a gap which should be proportional to the amount of disorder and/or additional localized features such as in \cite{Arean:2015wea,Arean:2015sqa}. Preliminary results indicate that this is indeed possible but just for large enough $\kappa$ such that the gradient instability discussed in Appendix \ref{cons} could appear. Moreover it would be interesting to check if we can accommodate for the formation of an hard gap connected with an exponential (and not a power law) suppression of the conductivity. Note that such a behaviour has been observed in similar models in \cite{Ling:2015epa, Kiritsis:2015oxa}.
Another interesting direction  relates to the possible instabilities at finite momentum present in our model for large enough $\kappa$ as introduced in Appendix \ref{cons}. This instability suggests that `ordered' (space-modulated) phases where homogeinity is lost should form at low temperatures, perhaps in a similar way as in the  previously discussed \textit{striped phases} and \textit{charge density wave excitations}  of \cite{Donos:2013gda,Donos:2013wia}. 
Note that strong disorder can indeed induce such  kind of situations in CM experiments and simulations \cite{exp1,exp2,exp3}.

Finally, it would be nice to understand better the relation  between S-Duality in the bulk (which is reflected into the particle-vortex duality of the CFT) and the presence of such a bound \eqref{bound} for the DC conductivity \cite{Grozdanov:2015qia}. It seemed that to produce an insulating behaviour or even a $\sigma_{DC}<1$ we had inevitably to spoil this duality (see \cite{Myers:2010pk} for previous discussions). According to our results it seems that this would not be the case, but a deeper study of this is necessary.

It is clear that the new coupling \eqref{direct} will have an impact also on many of the  properties of the material, such as for instance the (visco)elastic response (see \cite{Alberte:2015isw,Alberte:2016xja} for previous results). One expects a non-trivial mixing between electric and the mechanical response (`piezoelectric response') which would be interesting to study. For instance,  it is easy to see that the mass of the metric tensor mode is (\ref{MMM}) and therefore the elastic modulus and the viscosity must depend on the chemical potential \cite{Alberte:2015isw,Alberte:2016xja}. 
We leave these issues for future research.

\section*{Acknowledgements}
We thank M.Goykhman for useful comments about the draft. We thank Andrea Amoretti, Li Li, Wei-Jia Li, Yi Ling, Mikhail Goykhman, Saso Grozdanov, Elias Kiritsis, Alexander Krikun, Nicodemo Magnoli, Daniele Musso,  Koenraad Schalm,  Napat Poovuttikul for illuminating discussions and suggestions about the topic of this work. MB would like to thank University of Illinois, ICMT and P.Phillips for the warm hospitality during the completion of this note. We acknowledge support from MINECO under grant FPA2011-25948,  DURSI under grant 2014SGR1450 and Centro de Excelencia Severo Ochoa program, grant SEV-2012-0234. MB is supported by a PIF grant from Universitat Autonoma de Barcelona UAB.
\appendix
\section{Consistency}\label{cons}
In this appendix we discuss the consistency of the model \eqref{action}. We limit ourselves to the analysis in the decoupling limit where the metric is kept frozen. A complete analysis is of course interesting and needed but beyond the scope of this note. We checked rigorously the absence of any ghosty excitations and we made a preliminary analysis towards understanding possible gradient instabilities in the scalar sector.
\subsection{Transverse modes}
We define the transverse fluctuations as:
\begin{equation}
\delta A= A_T(t,u,y)\,dx\,\,\,\,\,\,,\,\,\,\,\,\,\delta \phi^x\,=\,\phi_T(t,u,y)
\end{equation}
In this sector the equations are decoupled.\\
The equation for the transverse part of the vector reads:
\begin{equation}
\partial_u\,A_T\, \left(\frac{f'}{f}+\frac{2 \,\alpha ^2\, u\, Y'\left(\alpha ^2 u^2\right)}{Y\left(\alpha ^2
   u^2\right)}\right)+\frac{\partial_y^2\,A_T}{f}-\frac{\partial_t^2\,A_T}{f^2}+\partial_u^2\,A_T\,=\,\frac{J_T}{2 \,u^2
   f\, Y\left(\alpha ^2 u^2\right)}
\end{equation}
where we have introduced a fictitious source for the transverse mode $\propto J_T A_T$ to keep track of the normalization of the kinetik term.\\
The speeds of sound are trivial but to avoid ghosty perturbations one needs:
\begin{equation}
\boxed{Y(X)>0}
\end{equation}
The equation for the transverse scalar reads:
\begin{align}
\frac{\partial_y^2\,\phi_T}{f}-\frac{\partial_t^2\,\phi_T}{f^2}+\chi \,\partial_u\,\phi_T+\partial_u^2\,\phi_T\,=\,\frac{J_{\phi}}{u^2 \,f\, \left(2 \,m^2\, V'\left(X\right)-\frac{\rho ^2\, u^4 \,Y'\left(X\right)}{Y\left(X\right)^2}\right)}
\end{align}
with $\chi$ a function of V and Y which we do not show explicitely and again a fake source $J_\phi$ to keep track of the normalization.\\
The speeds of sound are trivial but no ghosts in the transverse scalars implies:
\begin{equation}
\boxed{2 \,m^2\, V'\left(X\right)-\frac{\rho ^2\, u^4 \,Y'\left(X\right)}{Y\left(X\right)^2}\,>\,0}\label{c1}
\end{equation}
Notation maybe can be improved and made more physical using:
\begin{equation}
\bar{F}^2\,=\,-\frac{2\,u^4\,\rho^2}{\bar{Y}^2}
\end{equation}
such that condition \eqref{c1} reads:
\begin{equation}
\boxed{2 \,m^2\, V'\left(X\right)+\frac{\bar{F}^2}{2}\,Y'\left(X\right)\,>\,0}
\end{equation}
which is what we expect from the action.\\
If we want this bound to be satisfied at every charge density $\rho$ we need Y to have a negative slope.\\
The choice:
\begin{equation}
Y(z)\,=\,e^{-\kappa\,z}\,\,\,,\,\,\,\kappa>0
\end{equation}
we did in the main text satisfies indeed this requirement.\\
---------------------------------------------------------------------------------------------------------------
\\We can sum it up the constraints coming from the transverse sector as:
\begin{equation}
\boxed{V'(X)>0\,\,\,,\,\,\,Y(X)>0\,\,\,,\,\,\,Y'(X)<0}
\label{resum}
\end{equation}
Note how the last constraint implies that within this model disorder always decreases the DC electric conductivity as it happens in real condensed matter situations; this represents a wonderful match between theory and experiments.
\subsection{Longitudinal modes}
\begin{figure}
\centering
\includegraphics[width=7cm]{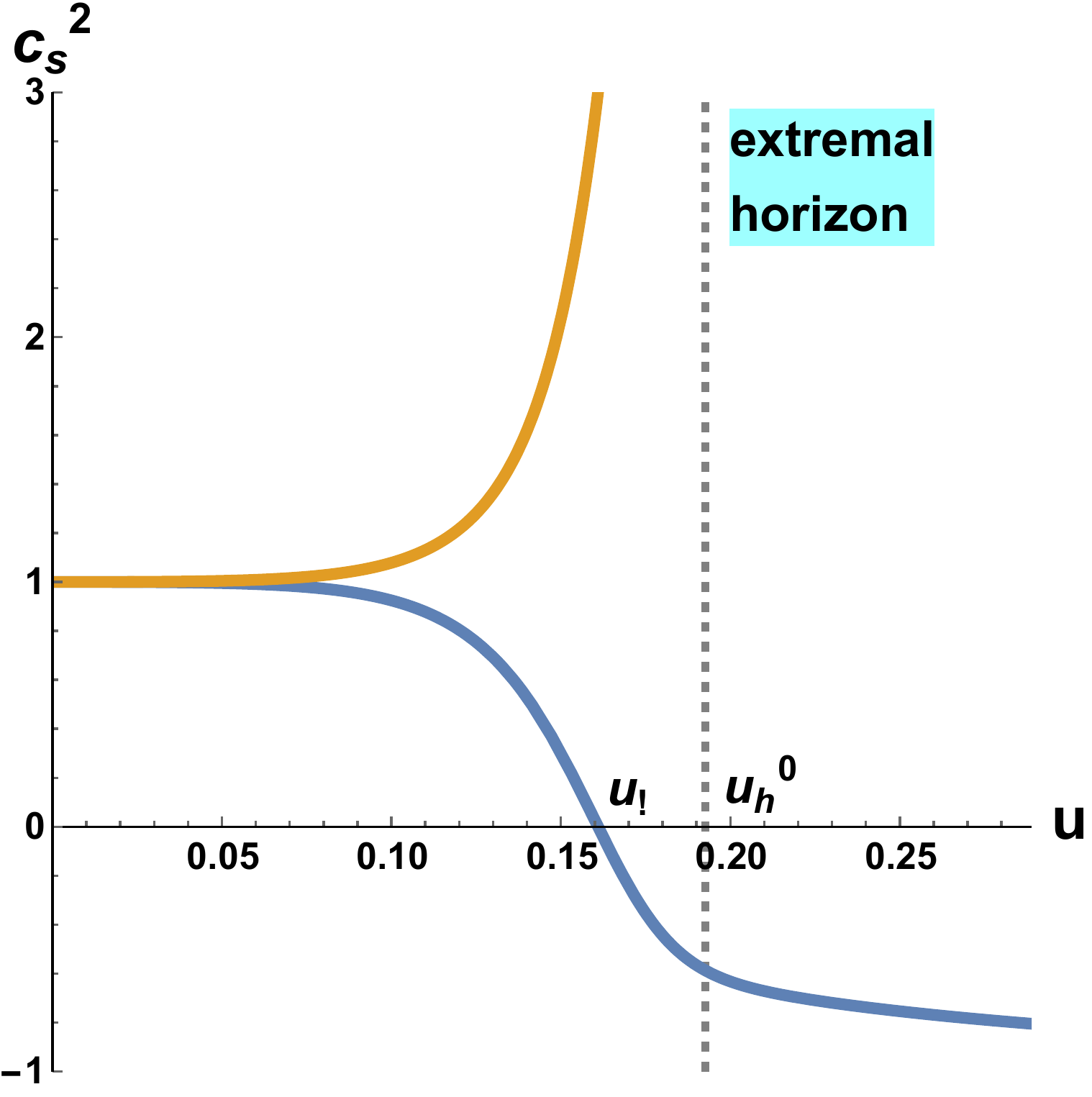}
\caption{Gradient instability for the scalar perturbations at $u=u_!$ for $\kappa=1,\rho=1,\alpha=10$ regarding the model considered in \eqref{themodel}.}
\label{instfig}
\end{figure}
\begin{figure}
\centering
\includegraphics[width=7cm]{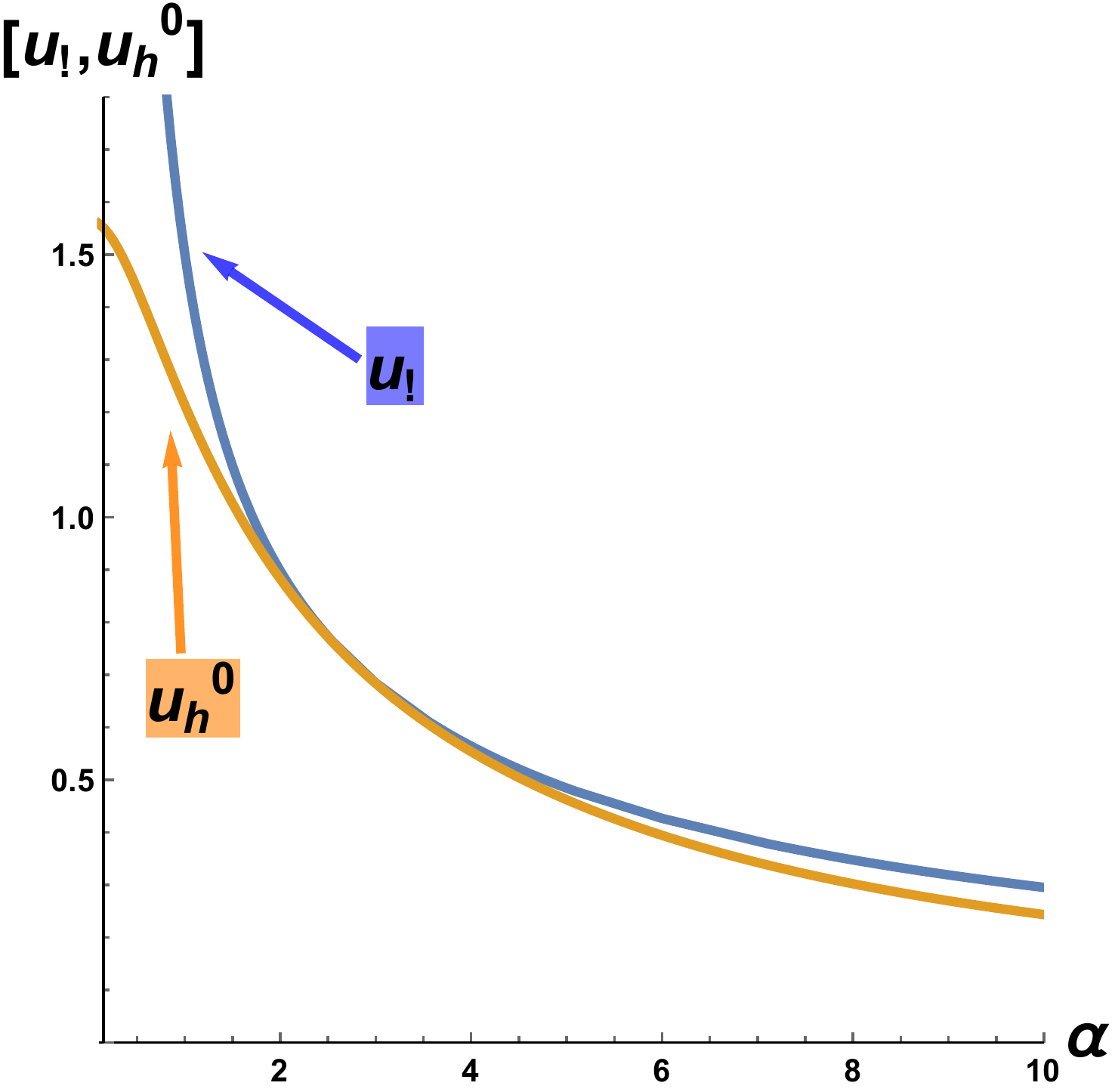}
\caption{No gradient instability for $\kappa=0.5,\rho=1$ upon varying the strength of disorder (i.e. graviton mass) for the model considered in \eqref{themodel}.}
\label{figsafe}
\end{figure}

We continue the analysis with the longitudinal modes of the perturbations:
\begin{align}
&\delta A= A_t(t,u,x)\,dt\,+\,A_u(t,u,x)\,du\,,\nonumber\\
&\delta \phi^x\,=\,\phi_L(t,u,x)
\end{align}
where we have already taken the gauge $A_x=A_y=0$.\\
We are left with three Maxwell equations which read:
\begin{align}
&\frac{f}{\bar{Y}}\,\partial_u\,\left[\bar{Y}\,\left(\partial_u A_t-\partial_t A_u\right)\right]\,-\,\frac{f}{\bar{Y}}\,\partial_u\,\left[\frac{u^2\,\alpha\,\rho\,\bar{Y}'}{\bar{Y}}\,\partial_x \phi_L\right]\,+\,\partial_x^2\,A_t\,=\,\frac{J_t}{2\,u^2\,\bar{Y}}\,\label{Forget}\\
&\frac{\partial_t A_t}{f}-\frac{1}{\bar{Y}}\,\partial_u\left(f\,\bar{Y}\,A_u\right)\,=\,\frac{J_x}{2\,u^2\,\bar{Y}}\,\label{Cons}\\
&\partial_x^2\,A_u\,+\,\frac{\partial_t\,\partial_u \,A_t}{f}-\frac{\partial_t^2 A_u}{f}-\frac{u^2\,\alpha\,\rho\,\bar{Y}'}{\bar{Y}^2\,f}\,\partial_t\,\partial_x \phi_L\,=\,\frac{J_u}{2\,u^2\,\bar{Y}}\label{DynA}
\end{align}
where $\bar{Y}=Y(u^2\,\alpha^2)$.\\
After some manipulations it is easy to see that the 3 equations are redundant and we can forget about the first one \eqref{Forget}. Note that one of the 2 equations remained (\ref{Cons}) is just a constraint such that there is just one dynamical longitudinal model inside the maxwell vector.\\
There is also a scalar equation left which reads:
\begin{align}
K\,\partial_x^2\,\phi_L\,+u^2\partial_u\left(\frac{f}{u^2}\,H\,\partial_u\,\phi_L\right)\,-\,\frac{H}{f}\,\partial_t^2\,\phi_L\,+\,\frac{2\,u^4\,\alpha\,\rho\,\bar{Y}'}{\bar{Y}}\,\left(\partial_u\,\partial_x\,A_t-\partial_t\,\partial_x\,A_u\right)\,=\,\frac{J_L}{u^2}\label{Dynphi}
\end{align}
with:
\begin{align*}
&H\,=\,2 \,m^2\, \bar{V}'-\frac{\rho ^2\, u^4\, \bar{Y}'}{\bar{Y}^2}\\
&K\,=\,2 \,\alpha ^2 \,m^2\, u^2\, \bar{V}''+2\,m^2\, \bar{V}'-\frac{\alpha ^2\, \rho ^2 \,u^6 \,\bar{Y}''}{\bar{Y}^2}-\frac{\rho ^2\, u^4\, \bar{Y}'}{\bar{Y}^2}
\end{align*}
Note that H has to be positive because of consistency in the transverse sector.\\
Now we have to reduce the equations (\ref{Cons}, \ref{DynA}, \ref{Dynphi}) to a coupled 2X2 system of equations for two variables using the constraint and then we can deal with the consistency of the longitudinal mixed sector.
For now on we suppress the sources and we go to Fourier space where it is easy to decouple the equations.\\
After a field redefinition $\phi_L \rightarrow \frac{k}{\omega}\phi_L $ one gets:
\begin{align}
&A_u''+\mathcal{C}_1\,A_u'\,+\, A_u\,\left(\frac{\omega ^2}{f^2}-\frac{k^2}{f}+\mathcal{C}_2\right)-\frac{\alpha \, k^2\, \rho\,  u^2\, \,
   \bar{Y}'}{f^2 \,\bar{Y}^2}\,\phi_L\,=\,0\,,\nonumber\\
   &\,\phi_L''+\frac{\mathcal{B}_1}{\mathcal{B}_3}\,\phi_L'+\phi_L
   \left(\frac{\omega ^2}{f^2}-\frac{k^2}{f}\,\frac{\mathcal{B}_2}{\mathcal{B}_3}\right)-\frac{2 \,\alpha  \,k^2\, \rho \, u^4 \, \bar{Y}'}{ \,\mathcal{B}_3\, \bar{Y}}\,A_u\,=\,0\,.
   \label{eqstab}
\end{align}
where $\{\mathcal{C}_1,\mathcal{C}_2,\mathcal{B}_1,\mathcal{B}_2,\mathcal{B}_3\}$ are non trivial functions of Y and V. In these variables the kinetik mixing is fully spacial (i.e. $\propto k^2$).\\
The functions $\mathcal{B}_{3}$ read:
\begin{equation}
\mathcal{B}_3\,=\,2 \,m^2\, \bar{V}'-\frac{\rho ^2 \,u^4\, \bar{Y}'}{\bar{Y}^2}
\label{b3expr}
\end{equation}
and it is strictly positive because of the requirements of the previous section summarized in \eqref{resum}. The other functions instead take the form:
\begin{align}
\mathcal{B}_1\,=\,&\frac{2 \,m^2\, f' \,V'}{f}+Y' \left(-\frac{\rho ^2 \,u^4 \,f'}{f Y^2}-\frac{2\, \rho ^2 \,u^3}{Y^2}\right)+4 \,\alpha ^2 \,m^2\,u\, V''-\frac{4 \,m^2\,
   V'}{u}-\frac{2 \,\alpha ^2 \,\rho ^2\, u^5\, Y''}{Y^2}\nonumber\\&+\frac{4 \,\alpha ^2\, \rho ^2\, u^5 \left(Y'\right)^2}{Y^3}\nonumber\\
\mathcal{B}_2\,=\,&2 \,\alpha ^2 \,m^2\, u^2 \,\bar{V}''+2 \,m^2\, \bar{V}'-\frac{\alpha ^2 \,\rho ^2\, u^6\, \bar{Y}''}{\bar{Y}^2}+\frac{2 \,\alpha ^2\, \rho ^2 \,u^6\, \bar{Y}'^2}{\bar{Y}^3}-\frac{\rho
   ^2\, u^4\, \bar{Y}'}{\bar{Y}^2}\,,\nonumber\\
    \mathcal{C}_1\,=\,&\frac{3\,f'}{f}+\frac{2\,u\,\alpha^2\,\bar{Y}'}{\bar{Y}}\nonumber\\
   \mathcal{C}_2\,=\,&\bar{Y}' \left(\frac{4 \,\alpha ^2\, u\, f'}{f\, \bar{Y}}-\frac{\alpha ^2 \,\rho ^2\, u^4}{f \,\bar{Y}^2}+\frac{2 \,\alpha ^2}{\bar{Y}}\right)+\frac{2\, f'}{f\,
   u}+\frac{\left(f'\right)^2}{f^2}+\frac{2 \,\alpha ^2 \,m^2\, \bar{V}'}{f}+\frac{2 \,\rho ^2\, u^2}{f \,\bar{Y}}+\nonumber\\+&\frac{4 \,\alpha ^4\, u^2\, \bar{Y}''}{\bar{Y}}-\frac{4\, \alpha ^4 \,u^2
   \left(\bar{Y}'\right)^2}{\bar{Y}^2}.
\end{align}
and they are a priori not positive definite.\\\\At this stage we could encounter several kind of instabilities (i.e. tachyons, gradient instabilities) but we can already state that there are no problem with ghosts. We will try in the following to at least scratch the surface of all these possible instabilities.\\ The first issue is that the kinetik matrix contains off diagonal terms proportional to $k^2$ which are going to affect the speed of sounds of the diagonal modes.\\
We are left with a u-dependent mixed kinetik matrix whose structure is:
\begin{equation}
\mathcal{K}\,=\,\begin{bmatrix}
\frac{\omega ^2}{f^2}-\frac{k^2}{f} & \frac{\alpha \, k^2\, \rho\,  u^2\, \,
   \bar{Y}'}{f^2 \,\bar{Y}^2} \\
-\frac{2 \,\alpha  \,k^2\, \rho \, u^4 \, \bar{Y}'}{ \,\mathcal{B}_3\, \bar{Y}} \,& \,\,\,\,\,\,\,\frac{\omega ^2}{f^2}-\frac{k^2}{f}\,\frac{\mathcal{B}_2}{\mathcal{B}_3}
\end{bmatrix}
\end{equation}
One can then proceed diagonalizing this matrix with a u-dependent rotation and extract the local speed of sound of the eigenmodes in the bulk\footnote{Note that a priori this quantity does not correspond to any physical speed of sound in the dual CFT but it deals essentialy with the stability of the bulk solution.} which reads:
\begin{equation}
c_{S\,\pm}^{2}\,=\,\,f\,\frac{\bar{Y}^{3/2} (\mathcal{B}_2 +\mathcal{B}_3)\pm\sqrt{\bar{Y}^3 (\mathcal{B}_3-\mathcal{B}_2)^2+8 \,\alpha ^2\, \mathcal{B}_3\, \rho ^2\, u^6\, \bar{Y}'^2}}{2 \,\mathcal{B}_3\,
   \bar{Y}^{3/2}}
   \label{speedtest}
\end{equation}
The first necessary but not sufficient stability check is having the speed of sound real valued.
This requirement can be then checked numerically starting from the expression \eqref{speedtest}.\\
For the model \eqref{themodel} considered in this work  the condition of no instability can be written down in a short form as:
\begin{equation}
e^{-\alpha ^2\, \kappa\,  u^2}+\kappa  \,\rho\, ^2 u^4 \,\left(1-\alpha ^2 \,\kappa \, u^2\right)\,>\,0\,.
\label{sign}
\end{equation}
For a generic choice of parameters and in particular for $\kappa$ being large we can encounter the appearance of a gradient instability signaled by an imaginary speed of sound or by expression \eqref{sign} getting negative at a certain value of the radial coordinate $u=u_!$; an example it is shown in fig.\ref{instfig}.
Whenever this mechanism happens before the extremal horizon position $u_h^0$ it represents a danger.\\
Through this note we will just consider points in the parameter space where no gradient instability appears (see fig.\ref{figsafe} for $\rho=1$) no matter the value of the disorder strength $\alpha$ and the charge density $\rho$ are. All the plots in the main text refer to the choice $\kappa=0.5$ which is indeed in the 'safe region' fig.\ref{figsafe}.
Note anyway that this could be not enough and the u-dependent rotation could introduce additional mass terms which could produce tachyonic instabilities; we will just introduce this issue in the next section.\\
To complete an exhaustive analysis of stability it would be great to study the QNM structure of the system to check if any of them could get a positive imaginary part signaling a concrete instability of the bulk background such as in \cite{Amado:2009ts}. This goes anyway beyond the scope of this short note.
\begin{figure}
\centering
\includegraphics[width=7cm]{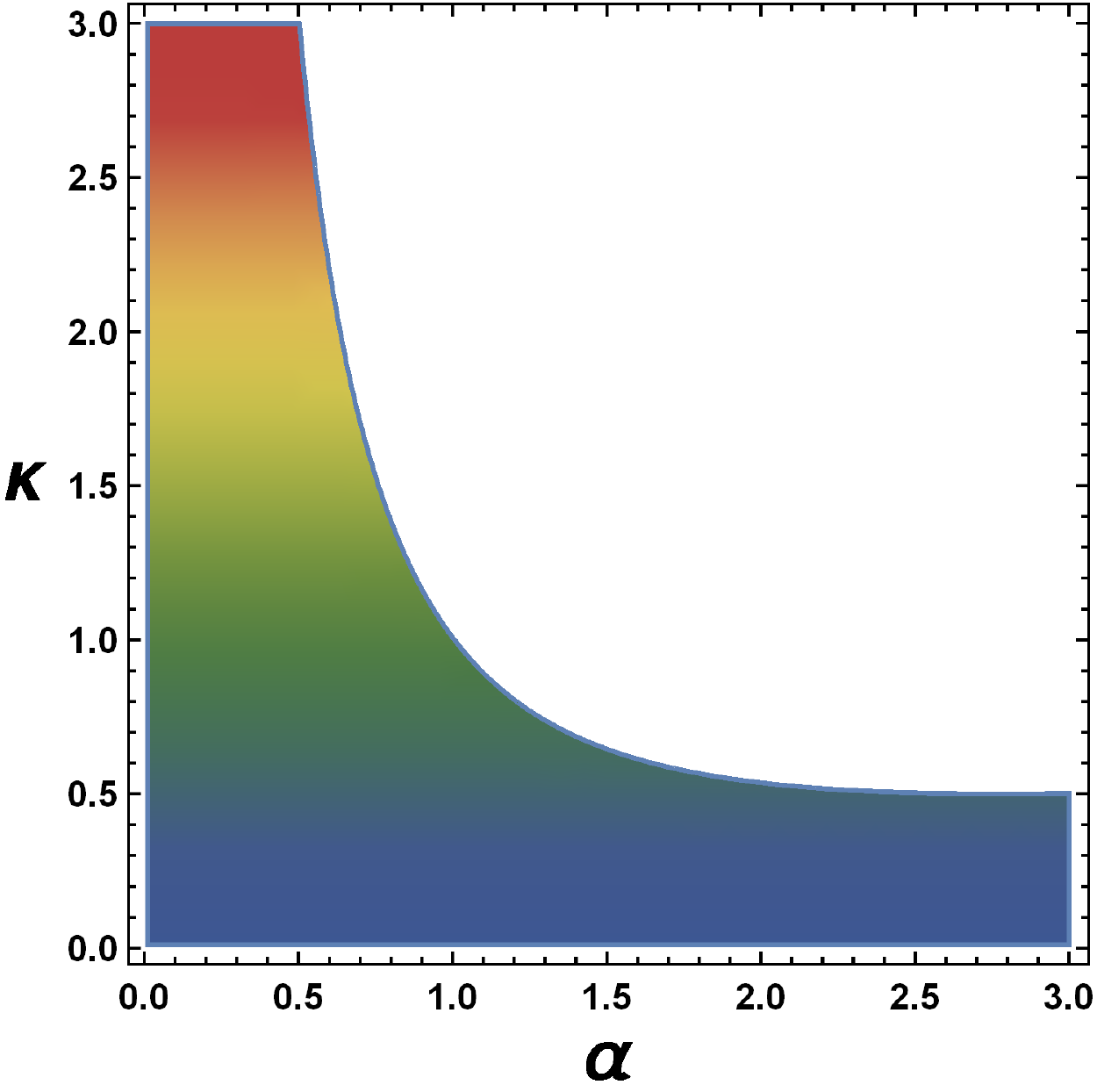}
\caption{Safe region for $\rho=1$ upon varying the strength of disorder (i.e. graviton mass) and the coupling $\kappa$ for the model considered in \eqref{themodel}.}
\label{cc}
\end{figure}
\subsection*{Possible disorder driven stripe instabilities}\label{strip}
Interestingly enough starting from equations \eqref{eqstab} it seems it could be possible to produce tachyonic instabilities at finite momentum. Technically speaking one can associate to equations \eqref{eqstab} a momentum dependent mass matrix:
\begin{equation}
\mathbb{M}_k\,=\,\begin{bmatrix}
-\mathcal{C}_2+\frac{k^2}{f} & -\frac{\alpha \, k^2\, \rho\,  u^2\, \,
   \bar{Y}'}{f^2 \,\bar{Y}^2} \\
\frac{2 \,\alpha  \,k^2\, \rho \, u^4 \, \bar{Y}'}{ \,\mathcal{B}_3\, \bar{Y}} \,& \,\,\,\,\,\,\,\frac{k^2}{f}\,\frac{\mathcal{B}_2}{\mathcal{B}_3}
\end{bmatrix}
\end{equation}
whose eigenvalues could violate the BF bound.\\
More specifically, this mechanism destabilizes the background at finite momentum (and at finite temperature) towards a different ground state where homogeinity may be lost and new interesting phases such as \textit{striped phases} may appear.\\
We leave the study of this possibility and its phenomenological consequences for the future.

\end{document}